\documentclass[twoside,slac_one]{revtex4}
\usepackage{graphicx}
\usepackage{fancyhdr}
\usepackage{amsmath} 
\usepackage{bm}
\usepackage{amsxtra}
\usepackage{amssymb}
\usepackage{amsthm}
\usepackage{latexsym}
\usepackage{lscape}

\begin{document}

\title{Identified Particle Production from the BES at RHIC}

%

\author{Lokesh Kumar (for the STAR collaboration)}
\affiliation{Physics Department, Kent State University, Kent, Ohio, USA}

\begin{abstract}
 The RHIC Beam Energy Scan focuses on the study of the QCD phase diagram --- temperature
 ($T$) vs. baryon chemical potential ($\mu_B$). The aim is to verify some predictions from QCD: that
 a cross-over occurs at $\mu_B$~=~0, that there exists
 a first-order phase transition at large $\mu_B$ and a
 critical point at an intermediate $\mu_B$. The spectra
 and ratios of produced particles can be used to extract
 $T$ and $\mu_B$ in different energies and system sizes.
 The STAR experiment has
 collected data for Au+Au collisions at $\sqrt{s_{NN}}=$
 7.7 GeV, 11.5 GeV, and 39 GeV in the year 2010. 
We present midrapidity $p_{T}$ spectra,
 rapidity density, average transverse mass, and particle
 ratios for identified hadrons from the STAR experiment.
 The centrality and transverse momentum dependence of the
 particle yields and ratios are compared to existing
 data at lower and higher beam energies.
The chemical and kinetic freeze-out conditions are extracted from the
ratios and particle spectra.
\end{abstract}

\maketitle

\thispagestyle{fancy}


\section{Introduction}
The RHIC beam energy scan (BES) is devoted 
to exploring the QCD phase diagram~\cite{rhicbes1,rhicbes2,qcdscience}. 
The QCD phase diagram represents the variation of temperature $T$
vs. baryon chemical potential $\mu_{\rm{B}}$.
At low temperatures, the relevant degrees of freedom are expected to
be hadronic but at high temperatures, the quarks and gluons are the 
relevant degrees of freedom of the system. In the QCD phase diagram, 
at $\mu_{\rm{B}}$ $\sim$ 0,
the transition from hadronic gas to quark gluon plasma (QGP) is expected to
be a crossover~\cite{lattice}. At large $\mu_{\rm{B}}$, it is expected
to be a first order phase transition~\cite{firstorder}. The point where the first order
phase transition line ends is called the QCD critical point which is
of current interest for many heavy-ion experiments
~\cite{rhicbes1,rhicbes2,cp_exp}. 
The main
objectives of RHIC BES are to search for the possible QCD phase
boundary
and to search for the possible QCD critical point in the QCD phase
diagram.
In addition, it will be interesting to see
how various observables, suggesting QGP formation at top RHIC
energies, behave as the collision energy is decreased. This
will help to locate the beam energy where there is no QGP formation or
the energy representing ``turn off'' of the QGP signatures.
A few potential observables for exploring the QCD phase diagram are: constituent-quark scaling of elliptic flow,
parity-violation in strong interactions, wiggle shape of directed flow
of protons, freeze-out eccentricity, $K/\pi$ ratio fluctuations, and
higher moments of net-protons~\cite{rhicbes1}.

The QCD phase diagram can be accessed by varying temperature $T$ and
baryonic chemical potential $\mu_{\rm{B}}$.
Experimentally this can be achieved by varying the colliding beam energy.
A $T$-$\mu_{\rm{B}}$ space point in the QCD phase diagram can be
obtained from the spectra and ratios of the produced hadrons. Once the
$T$-$\mu_{\rm{B}}$ space point is obtained, one can study various 
signatures, mentioned above, for evidence of the possible QCD phase boundary and QCD critical
point. 
In the year 2010, the
STAR experiment collected data for Au+Au collisions at energies $\sqrt{s_{NN}}=$ 7.7,
11.5, and 39 GeV, as part of the proposed energies for the BES program,
in addition to usual energies of 62.4 and 200 GeV.

\section{Data Analysis}
The data presented here are
for Au+Au collisions at $\sqrt{s_{NN}}$ = 7.7, 11.5, and 39 GeV taken
in 2010 by the STAR experiment~\cite{star}. 
The total events analyzed for the results presented here are about 4 M, 8 M, and 10 M, respectively for $\sqrt{s_{NN}}$ = 7.7, 11.5, and 39 GeV.
The trigger
selection is done by using the Beam Beam Counter (BBC) and the Vertex Position Detector (VPD)~\cite{vpd}.
The main subsystem used for particle identification is the Time Projection Chamber (TPC)~\cite{tpc}.
The particle identification is enhanced to higher $p_{T}$ with the
inclusion of the full barrel
Time Of Flight (TOF) detector~\cite{tof}.
The centrality selection is done using the uncorrected charged track
multiplicity
measured event-wise in the TPC within $|\eta|<$ 0.5. The centrality classes represent the
fractions of this multiplicity distribution.  
The average number of 
participating nucleons ($\langle N_{\rm{part}} \rangle$) and 
collisions  ($\langle N_{\rm{coll}} \rangle$)  are obtained
by comparing the multiplicity distribution with that from Glauber
Monte Carlo simulation~\cite{rhicbes2}.
The raw yields are extracted
at low $p_{T}$ using ionization energy loss ($dE/dx$) from TPC, and 
using TOF information at higher $p_{T}$. 
The raw spectra are corrected for the detector acceptance and tracking
efficiency. These correction factors are obtained 
together by embedding the tracks simulated using the GEANT model of
the STAR detector into real events at the raw data level~\cite{geant}.
The results are presented for midrapidity $|y|<0.1$
region.

\section{Results and Discussions}

\subsection{Transverse Momentum Spectra}
\begin{figure}
\begin{center}
\includegraphics[scale=0.33]{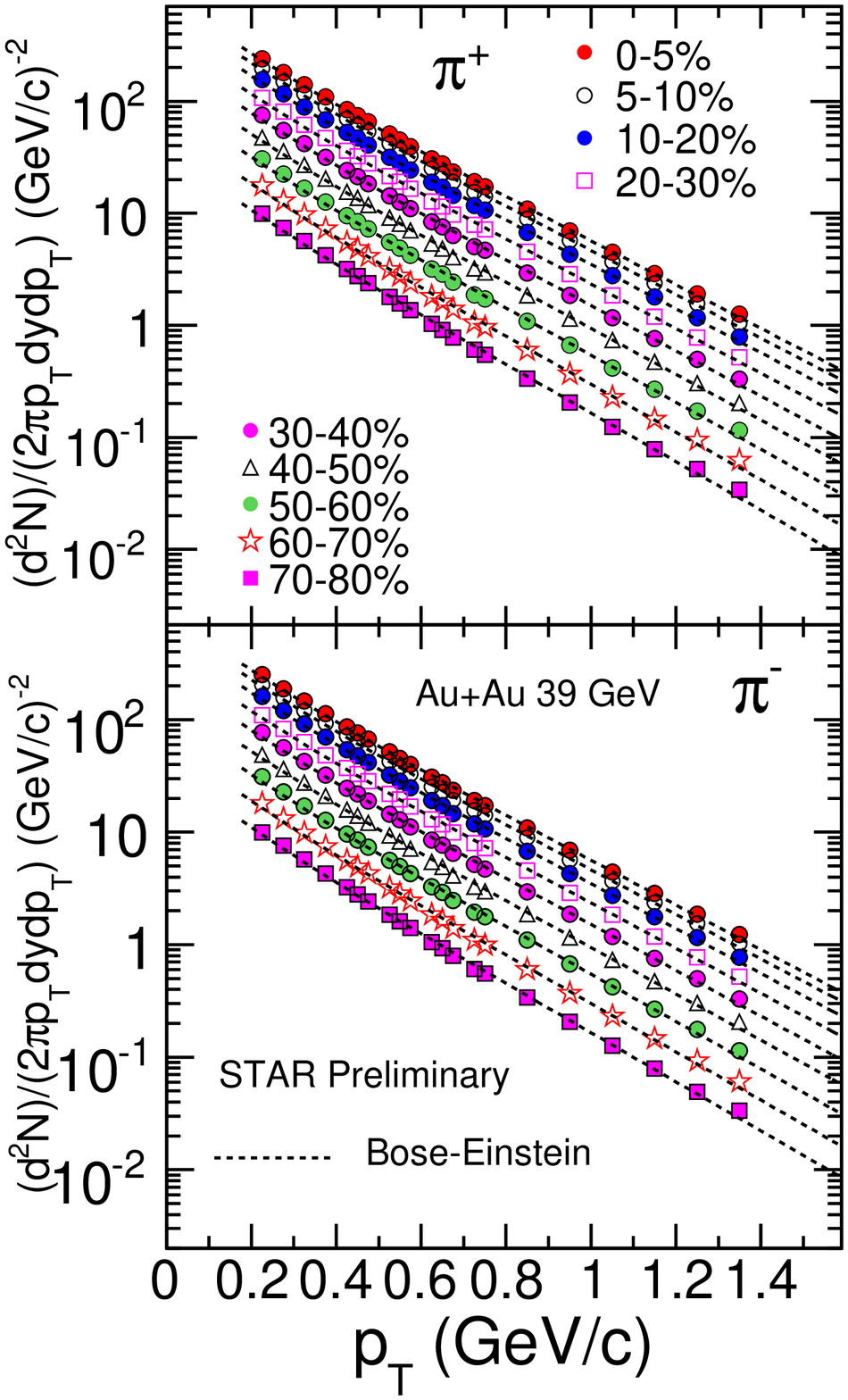}
\includegraphics[scale=0.33]{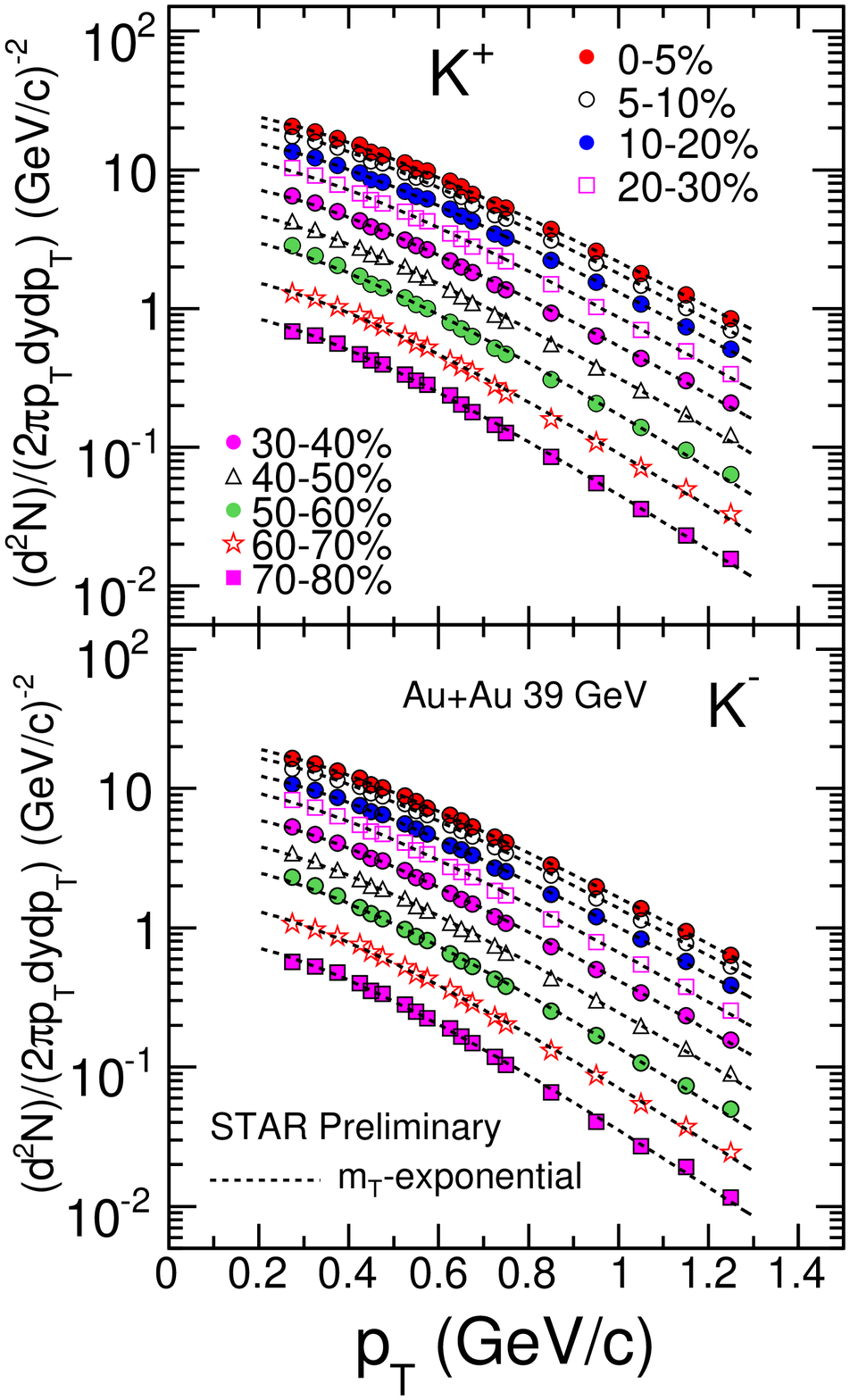}
\caption{\label{spectrafig} 
Transverse momentum spectra for charged pions $\pi^{\pm}$ (left panel) and charged
kaons $K^{\pm}$ (right panel) for various collision centralities at midrapidity ($|y|<$
0.1) in Au+Au collisions at $\sqrt{s_{NN}}=$ 39 GeV.  Errors shown are statistical.
}
\end{center}
\end{figure}

Figure~\ref{spectrafig} shows invariant yields $d^2N/(2\pi p_Tdydp_T$)
vs. transverse momentum $p_T$ for $\pi^{\pm}$ (left panel)
$K^{\pm}$ (right panel) and for various collision centralities at midrapidity ($|y|<$
0.1) in Au+Au collisions at $\sqrt{s_{NN}}=$~39~GeV.  The curves
represent fits to the spectra: pion spectra are fitted with Bose-Einstein function
and the kaon spectra with $m_T$-exponential. 
The $dN/dy$ and $\langle p_{T} \rangle$ or $\langle m_{T} \rangle$ values
are obtained using the data in the measured $p_{T}$ ranges and
extrapolating using a fit function (Bose-Einstein for $\pi$ and $m_{T}$-exponential
for $K$) for the unmeasured $p_{T}$ ranges. The contribution to the
yields from extrapolation to the total yield is about 20-30$\%$.
The pion spectra presented here
are corrected for weak-decay feed-down and muon contamination using
STAR HIJING+GEANT, as was done for previous STAR
results~\cite{starpid}. 
The total contribution from weak-decay feed-down and muon
contamination is about 17\% at low $p_{T}$
and becomes almost negligible ($\sim$ 1\%) around $p_T$~=~1.4~GeV/$c$.

\subsection{Centrality Dependence of Yields and Average Transverse Momentum}
\begin{figure}
\includegraphics[scale=0.4]{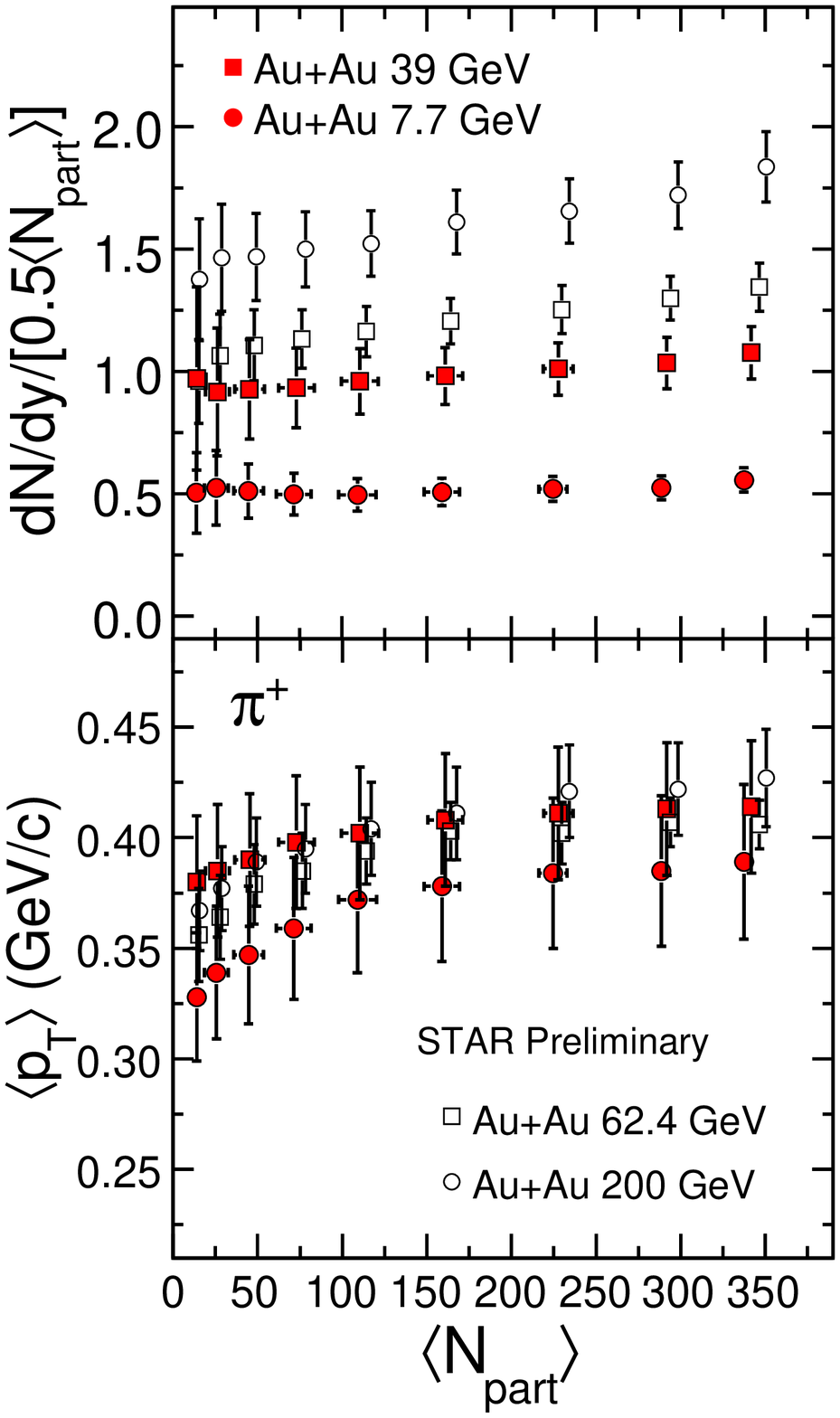}
\includegraphics[scale=0.4]{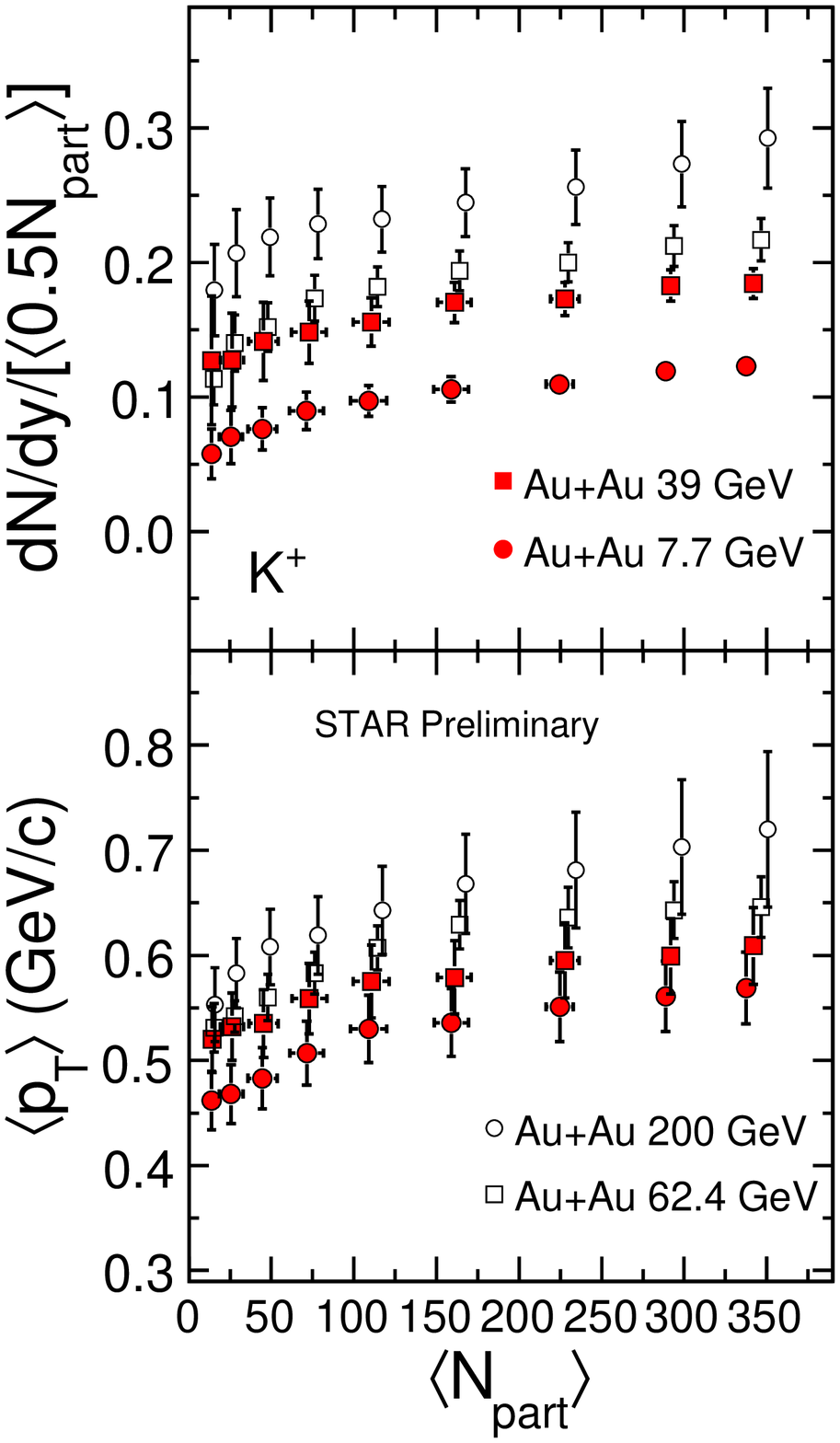}
\caption{\label{centdep} 
Left: Variation of $dN/dy$ divided by participating
  nucleon pairs (top panel) and $\langle p_{T} \rangle$ (bottom panel)
with collision centrality for $\pi^{\pm}$. Results from
  7.7 and 39 GeV BES data are compared with those from previous STAR
  measurements~\cite{starpid}. 
  Errors are statistical and systematic added in quadrature. Right:
  Similar measurements for  $K^{\pm}$.  
}
\end{figure}

Figure~\ref{centdep} shows the centrality dependence of $dN/dy$
normalized by number of participating nucleon pairs 
$\langle N_{\rm{part}}\rangle / 2$ (top panels)  and $\langle p_{T}
\rangle$  (bottom panels) for
$\pi^{\pm}$ (left) and for $K^{\pm}$ (right) in Au+Au collisions at $\sqrt{s_{NN}}=$ 7.7 and 39 GeV. New
measurements from BES data are compared to previously
published results at top RHIC energies~\cite{starpid}. The $dN/dy$ per
participating nucleon pair for pions is almost constant as a function of
collision centrality at $\sqrt{s_{NN}}=$ 7.7 GeV, while it
increases with $N_{\rm{part}}$ for higher energies and for kaons.
The $\langle p_{T} \rangle$ increases with collision centrality for
both pions and kaons at all
energies, indicating that the average
collective velocity in the radial direction increases with collision centrality.

\subsection{Energy Dependence of Yields and Average Transverse Mass}
\begin{figure}
\begin{center}
\includegraphics[scale=0.33]{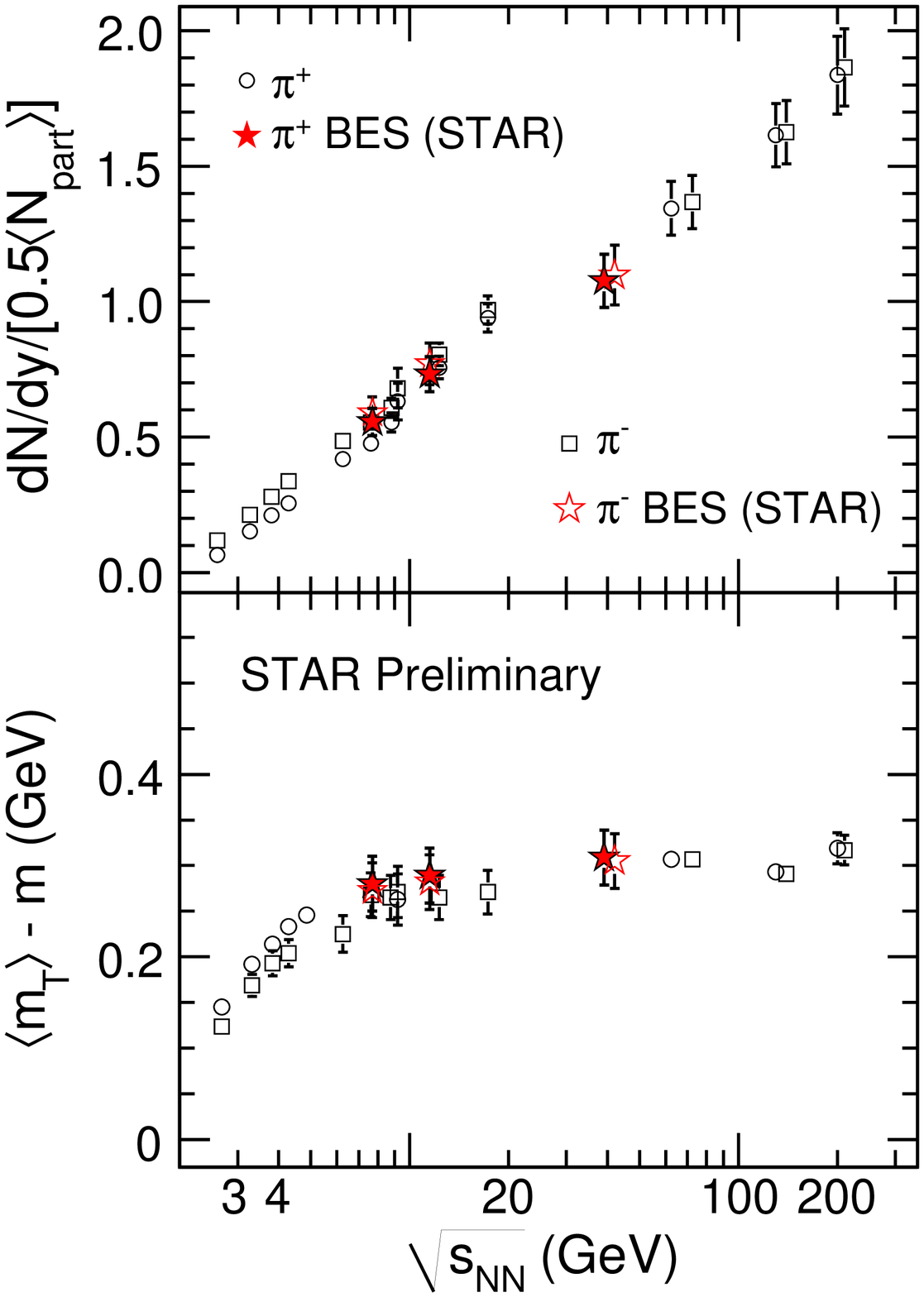}
\includegraphics[scale=0.33]{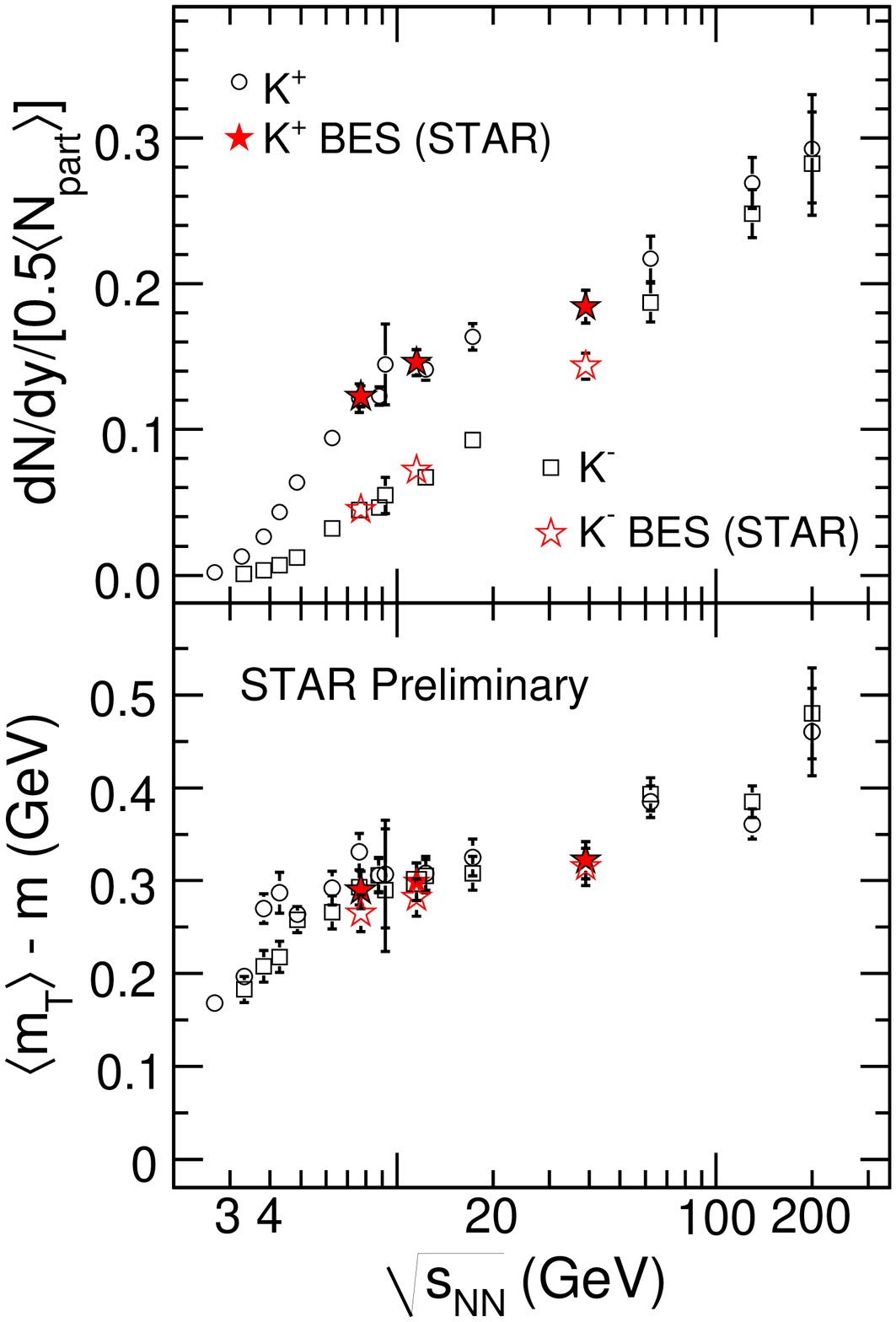}
\caption{\label{edep} Left: $dN/dy$ divided by participating
  nucleon pair (top panel) and $\langle m_{T} \rangle - m$ (bottom panel),
  plotted as a function of beam energy for $\pi^{\pm}$. Results from
  BES data are compared with those from previous
  measurements~\cite{starpid,sps,ags}. 
  Errors are statistical and systematic added in quadrature. Right:
  Similar measurements for  $K^{\pm}$.  
}
\end{center}
\end{figure}
\begin{figure}
\includegraphics[scale=0.37]{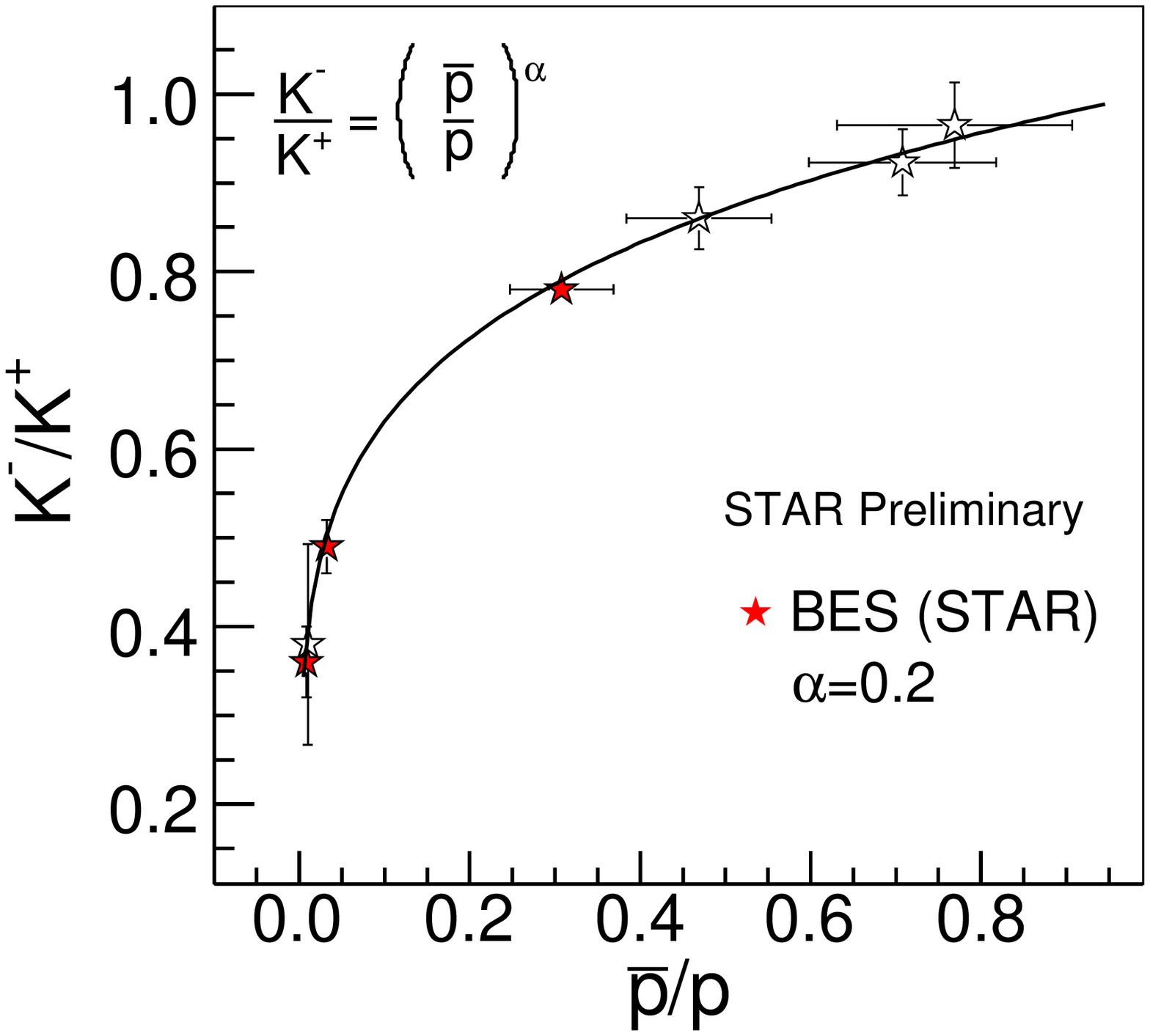}
\includegraphics[scale=0.35]{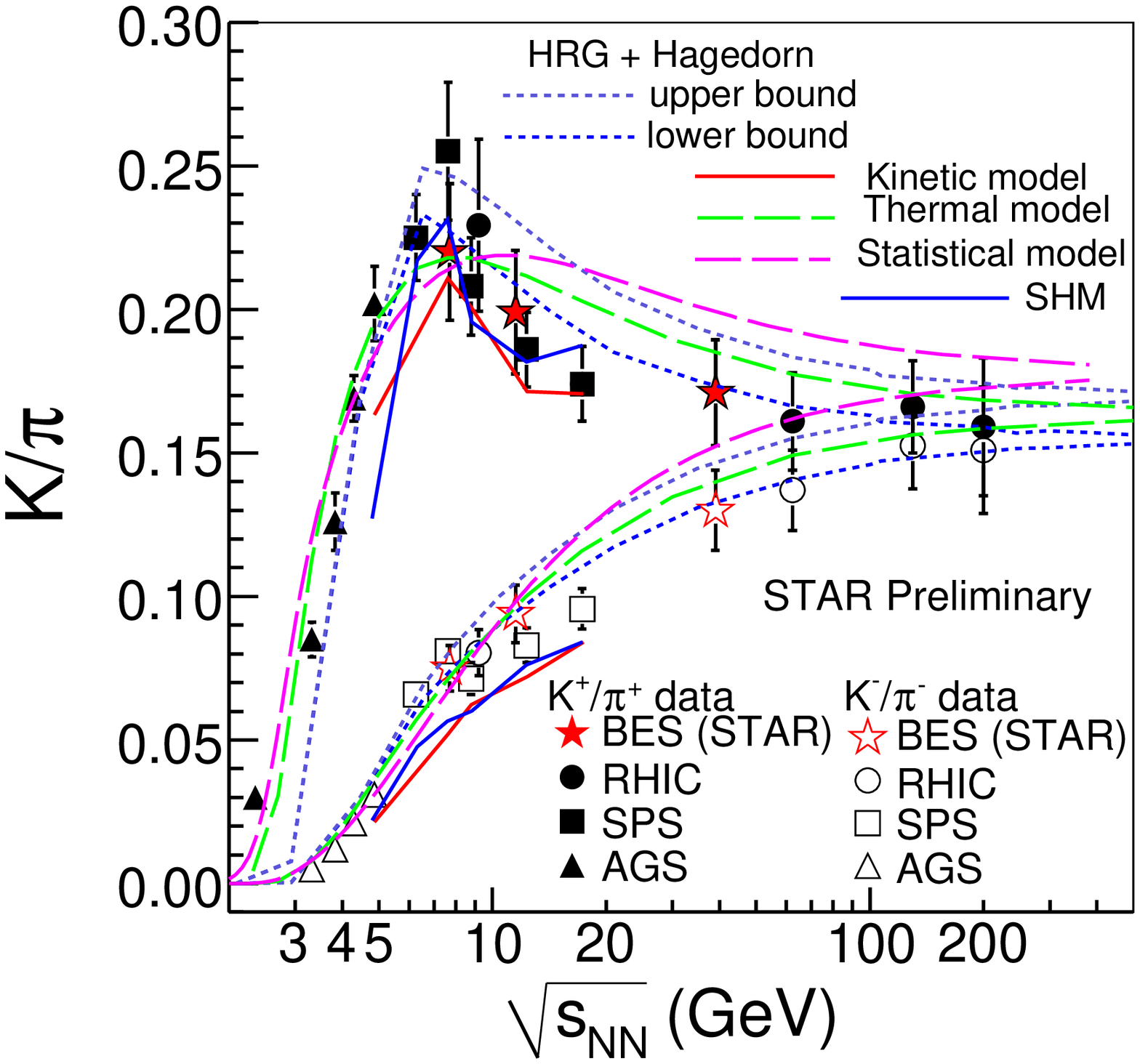}
\caption{\label{ratios}
 Left: Correlation of $K^{-}/K^{+}$ ratio with $\bar{p}/p$
ratio for central collisions at midrapidity, shown for new measurements from BES data and for
the published results from RHIC energies~\cite{starpid}. The curve
represents a
the power law fit to the data. 
Errors are statistical and systematic added in quadrature.
 Right: Energy dependence of $K^{\pm}/\pi^{\pm}$
  ratio for central collisions at midrapidity. New results from BES
  data are compared to previously published
  results at RHIC~\cite{starpid}, SPS~\cite{sps}, and AGS~\cite{ags}. Errors are statistical and systematic
  added in quadrature. Results are also compared with various
 theoretical model predictions~\cite{shm,stat,therm,kin,hrghag}. 
}
\end{figure}
\begin{figure}
\begin{center}
\includegraphics[scale=0.35]{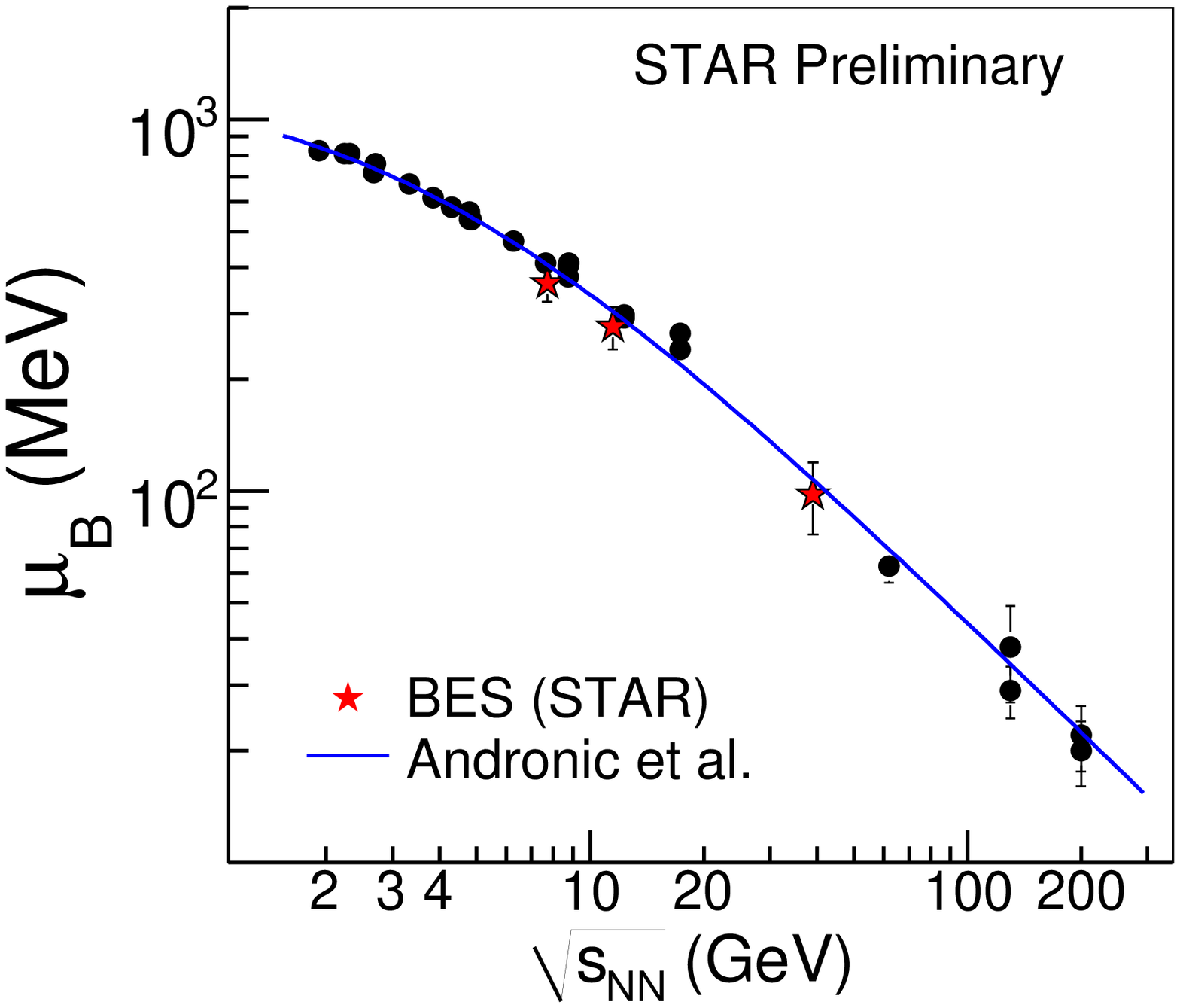}
\includegraphics[scale=0.37]{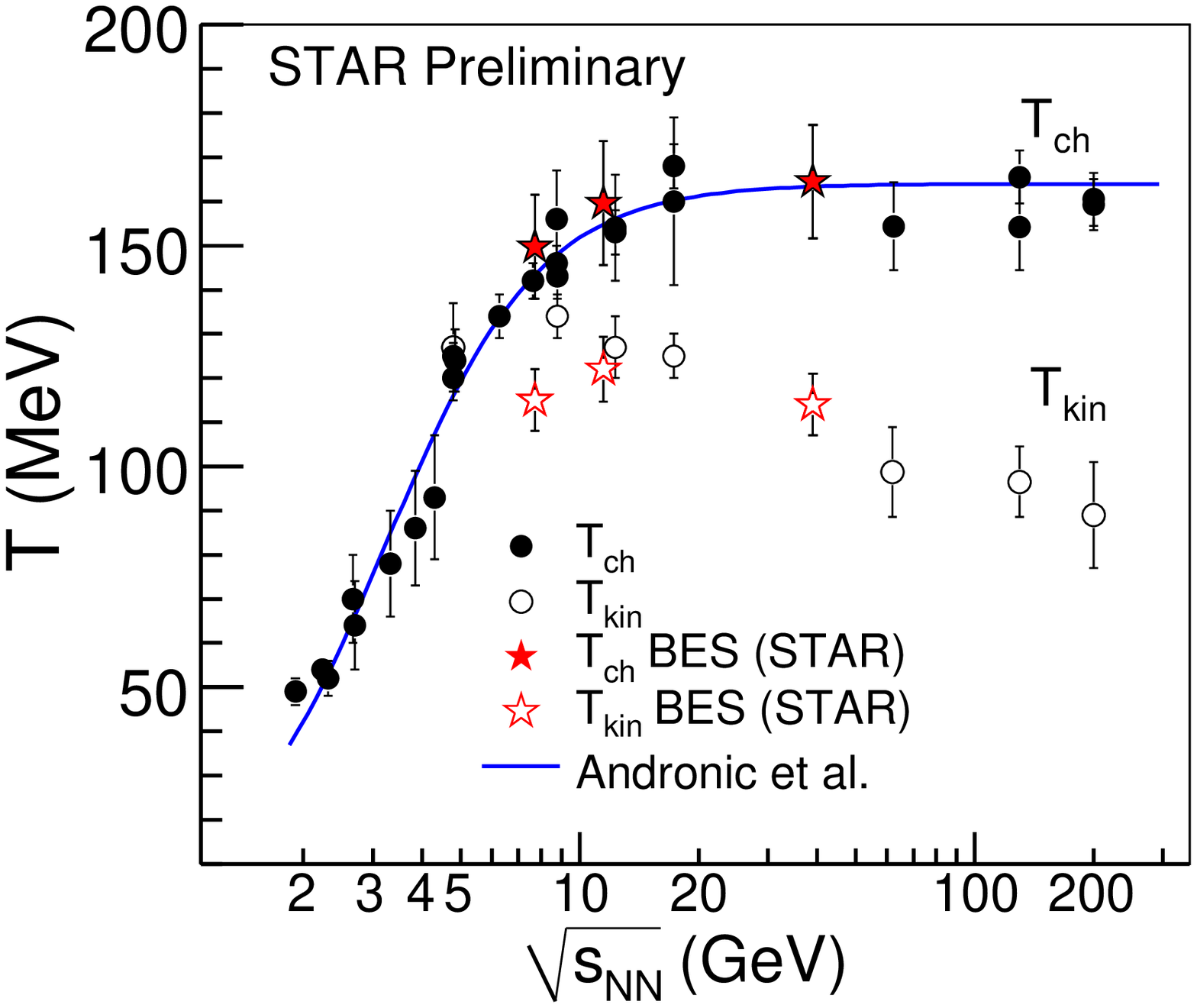}
\includegraphics[scale=0.35]{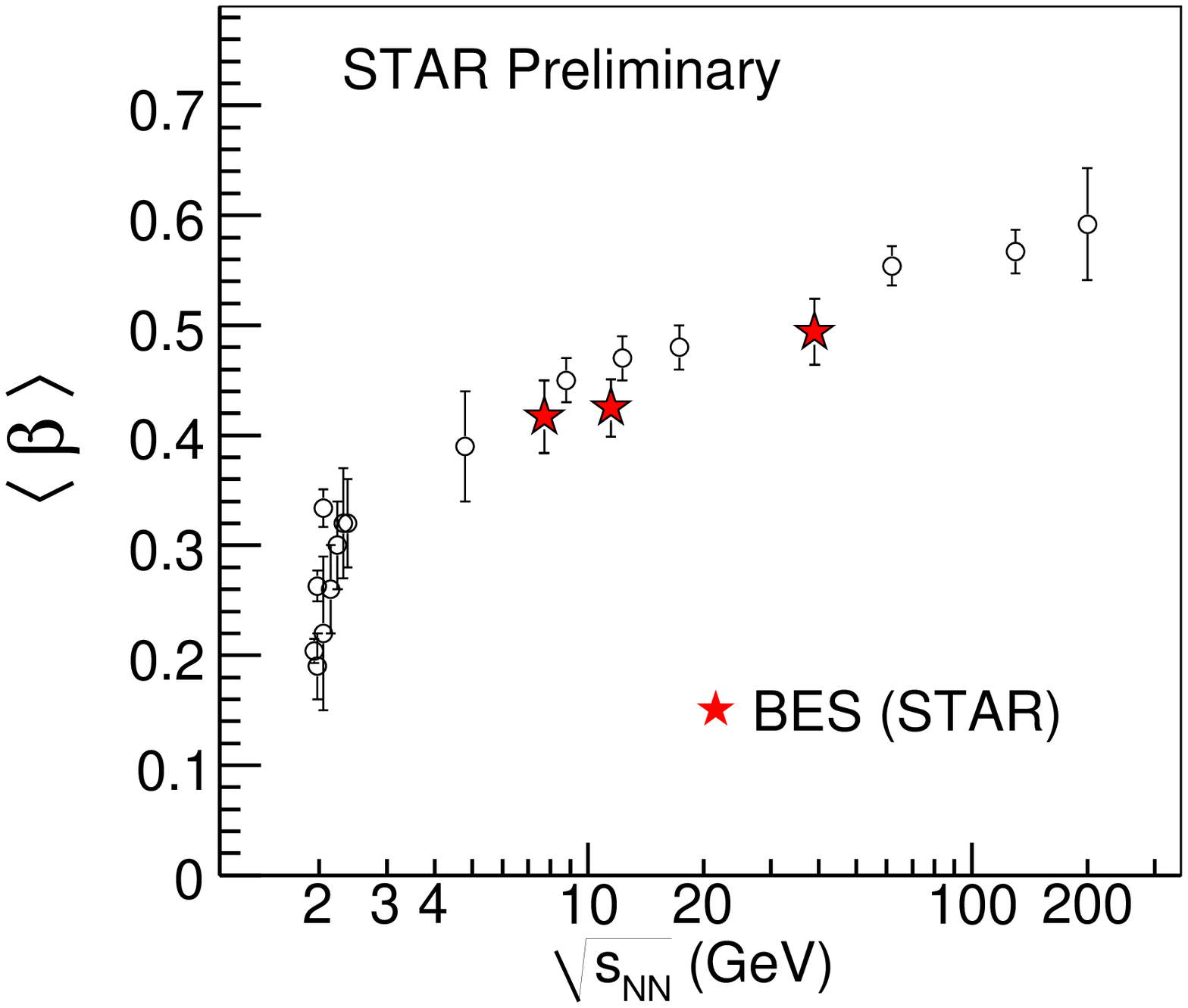}
\includegraphics[scale=0.35]{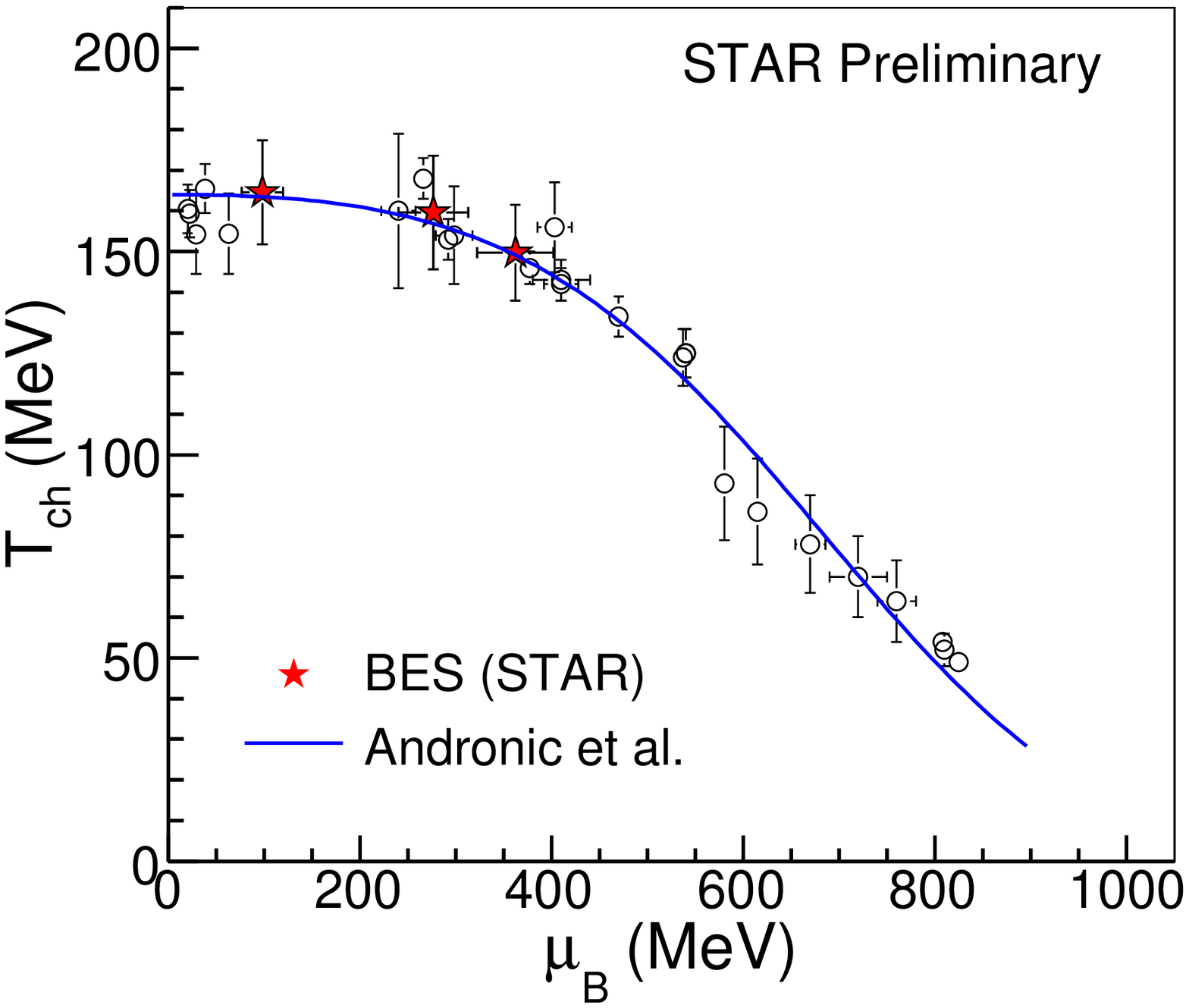}
\caption{\label{freezeout} Energy dependence of
  baryonic chemical potential (top-left), temperature
 ($T_{\rm{ch}}$, $T_{\rm{kin}}$) (top-right),
 and average flow velocity (bottom-left).  
Bottom-right: Chemical freeze-out temperature
 plotted as a function of baryonic chemical potential. New results
 from BES data are shown with star symbols while other results are
 from the Ref.~\cite{starpid} and references therein.
 Curves represent the parametrization from Ref.~\cite{anton}.
}
\end{center}
\end{figure}

Figure~\ref{edep} shows the energy dependence of $dN/dy$ normalized by
$\langle N_{\rm{part}}\rangle / 2$ (top panels)  and $\langle m_{T}
\rangle - m $  (bottom panels) as a function of $\sqrt{s_{NN}}$ for
$\pi^{\pm}$ (left) and for $K^{\pm}$ (right). The star symbols
represent the data from BES energies and are compared with previous
measurements~\cite{starpid,sps,ags}. The results are shown for the
central collisions and for midrapidity regions. The STAR BES results are consistent with the
published beam energy dependence trend. The yields per participating
nucleon pair
increase with beam energy. 
The quantity $\langle m_{T} \rangle - m$, where $m_{T}
=\sqrt{p_{T}^{2} + m^{2}}$ and $m$ is the hadron mass, increases
with beam energy for lower energies, becomes almost constant for the region
covered by the BES data and then tends to increase towards the top RHIC
energies. This is an interesting observation in a scenario where the
system is in a 
thermodynamic state. In that case, $\langle m_{T} \rangle - m$ can be
related to temperature of the system and $dN/dy~(\propto
\log(\sqrt{s_{NN}}))$ may represent
entropy. Then this observation could reflect the
signature of a first order phase transition as proposed in the
Ref.~\cite{vanhove}. However, 
other interpretations of the observed
$\langle m_{T} \rangle - m$ are possible~\cite{bedanga}.

\subsection{Particle Ratios}
The particle ratios provide useful information about the collision
dynamics as these are fixed after the chemical
freeze-out (see Sec. 3.5). Figure~\ref{ratios} (left panel) shows the correlation of
$K^{-}/K^{+}$ ratio with $\bar{p}/p$ ratio for new measurements from
BES energies along with our previously published
results~\cite{starpid}. This could give information on how the kaon
production is related to the net-baryon density which is reflected by the
$\bar{p}/p$ ratio.  
The STAR $p$ and $\bar{p}$ yields are not feed-down corrected.
At lower energies, the kaon production is
dominated by the associated production which results in more $K^{+}$
production compared to $K^{-}$. Also the $\bar{p}/p$ ratio is much
less than
unity at midrapidity, indicating that there is large baryon stopping
at midrapidity at the lower energies.
As we go towards higher energies, the pair production mechanism
starts to dominate and the ratios tend to become closer to
unity. The
correlation between $K^{-}/K^{+}$ (representing net-strange chemical
potential, $\mu_{S}$) and $\bar{p}/p$ (representing net-baryon
chemical potential, $\mu_{B}$) ratios seem to
follow a power law behavior with $\alpha$=0.2 (represented by the
curve). In a hadron gas, the
relationship between $\mu_{S}$ and $\mu_{B}$ depends on the
temperature. In a particular case of $T$=190 MeV and $\mu_{B} <$ 500
MeV, these potentials follow the relation 
$\mu_{S} =(1/3) \mu_{B}$~\cite{kvsp}.

Recent theoretical calculations~\cite{cleyman2} suggest that the
maximum net-baryon density at freeze-out is attained at the lowest BES
energy of $\sqrt{s_{NN}} \sim $ 7.7 GeV. The maximum net-baryon density 
could also be related to the peak observed in the energy dependence of $K^{+}/\pi^{+}$
ratio at around $\sqrt{s_{NN}} \sim $ 7-8 GeV, as was observed by the
NA49 experiment~\cite{sps}. 
This is sometimes referred to as the ``horn''.
 The $K/\pi$ ratio is also important as it could suggest 
the strangeness enhancement in heavy-ion collisions with respect to
the elementary collisions. Figure~\ref{ratios} (right panel) shows the
energy dependence of $K^{\pm}/\pi^{\pm}$ ratio for central collisions
at midrapidity. The BES results are
shown with star symbols and are compared
to results from previous measurements~\cite{starpid,sps,ags}. 
The BES results are in good agreement with the trend of energy dependence established by the published measurements. 

Figure~\ref{ratios} (right panel) also shows the predictions of energy dependence of
$K/\pi$ ratio from various theoretical model
calculations. 
The energy dependence of $K^{+}/\pi^{+}$ ratio has been
interpreted using the Statistical Model of Early Stage
(SMES)~\cite{smes}. The model predicts first order phase transition
and the existence of mixed phase around beam energy of 7-8 GeV.
The SHM or Statistical Hadronization Model~\cite{shm} assumes that the strong
interactions saturate the particle production matrix elements. 
This means that the yield of particles is controlled predominantly by the
magnitude of the accessible phase space. 
The system is in chemical non-equilibrium for 
$\sqrt{s_{NN}}<$ 7.6 GeV, while for higher energies, 
the over-saturation of chemical occupancies is observed.
  The Statistical Model~\cite{stat} assumes that the ratio
  of entropy to $T^3$ as a function of collision energy increases for
  mesons and decreases for baryons. Thus, a rapid change is expected
  at the crossing of the two curves, as the hadronic gas undergoes a
  transition from a baryon-dominated to a meson-dominated gas. The
  transition point is characterized by $T$=140 MeV, $\mu_B$= 410 MeV,
  and $\sqrt{s_{NN}}$=8.2 GeV. 
In the Thermal Model~\cite{therm}, the energy
  dependence of $K^{\pm}/\pi^{\pm}$ is studied by including
$\sigma$-meson, which is neglected in most
  of the models, and many higher
  mass resonances ($m>$ 2 GeV/$c^2$) into the resonance spectrum employed in
  the statistical model calculations. 
  The hadronic non-equilibrium Kinetic model~\cite{kin} assumes
  that the surplus of strange particles is produced in secondary
  reactions of hadrons generated in nuclear collisions. Then the two
  important aspects are the available energy density and the lifetime
  of the fireball.
It is suggested that these two aspects combine in such a way to show a sharp peak for
the strangeness-to-entropy or $K/\pi$ ratio as a function of beam energy.
In the Hadron Resonance Gas and Hagedorn model (HRG+Hagedorn)~\cite{hrghag}, all hadrons
  as given in PDG  with masses up to 2 GeV$c^2$ are included. The unknown
  hadron resonances in this model are included through Hagedorn's
  formula for the density of states. The model assumes that the strangeness in the baryon sector
  decays to strange baryons and does not contribute to the kaon
  production.
The energy dependence of $K^{\pm}/\pi^{\pm}$ ratio
seems to be best explained using HRG+Hagedorn model.

\subsection{Freeze-out Conditions}
The integrated yields of hadrons change only through the inelastic
collisions. The point in time when these inelastic collisions stop is
called the chemical freeze-out. After this stage, the particle ratios are
frozen and the system can be described by the thermal equilibrium
model~\cite{starpid,chemfo} which involves two main parameters $T_{\rm{ch}}$ and
$\mu_{B}$.  The $T_{\rm{ch}}$  and $\mu_{B}$ are obtained by fitting the
particle ratios with the thermal model. 
The point in time when the 
the elastic collisions among the particles cease is called the kinetic freeze-out.
The system can be described by the blast-wave (BW)
formulation involving two main parameters $T_{\rm{kin}}$ and average
flow velocity ($\langle \beta \rangle$)~\cite{starpid,bw}. The $T_{\rm{kin}}$ and $\langle
\beta \rangle$ are obtained by simultaneously fitting the invariant yields for $\pi,
K$, and $p$ with the blast-wave model.

Figure~\ref{freezeout}  shows the variation of extracted freeze-out
parameters for central collisions. 
Top-left plot shows the energy dependence of baryonic chemical
potential. It shows that $\mu_{B}$ decreases as the beam energy
increases. This is expected
since there are fewer net-baryons at midrapidity at higher energies
because of less stopping of baryons at midrapidity at higher
energies. Top-right plot shows the kinetic freeze-out (open symbols) and
chemical freeze-out (solid symbols) temperature plotted
vs. $\sqrt{s_{NN}}$. The chemical freeze-out temperature increases with energy
and saturates at the higher energies, and kinetic freeze-out
temperature decreases with beam energy after $\sqrt{s_{NN}} \sim$ 7.7
GeV. Bottom-left plot shows the average flow velocity
plotted vs. $\sqrt{s_{NN}}$. The $\langle \beta \rangle$ increases
with beam energy.
Bottom-right plot shows the current picture of the phase diagram ($T_{\rm{ch}}$
vs. $\mu_{B}$). The new measurements from BES program now extend the
$\mu_{B}$ range covered by the RHIC from 20--400 MeV. 

\section{Summary}

In summary, bulk properties from the RHIC BES program are presented in
this paper. The BES data are nicely fitted in the beam energy
dependence trend established previously. The yields of identified hadrons increase with beam
energy. The $\langle m_{T} \rangle - m$ is almost constant for the BES
energies. The $\langle p_{T} \rangle$ increases with centrality
reflecting collectivity increases with centrality. The net-baryon density plays an important role at the lower
energies as seen in the energy dependence of $K/\pi$ ratio, and the
correlation between $K^{-}/K^{+}$ and $\bar{p}/p$ ratios. The new
measurements from the BES program extend the $\mu_{B}$ range covered by the RHIC from 20--400 MeV.

\begin{acknowledgments}
We thank the RHIC Operations Group and RCF at BNL, the NERSC Center at
LBNL and the Open Science Grid consortium for providing resources and
support. This work was supported in part by the Offices of NP and HEP
within the U.S. DOE Office of Science, the U.S. NSF, the Sloan
Foundation, the DFG cluster of excellence `Origin and Structure of the
Universe' of Germany, CNRS/IN2P3, FAPESP CNPq of Brazil, Ministry of Ed. and Sci. of the Russian Federation, NNSFC, CAS, MoST, and MoE of China, GA and MSMT of the Czech Republic, FOM and NWO of the Netherlands, DAE, DST, and CSIR of India, Polish Ministry of Sci.\ and Higher Ed.\, Korea Research Foundation, Ministry of Sci.\, Ed.\ and Sports of the Rep.\ Of Croatia, and RosAtom of Russia.

\end{acknowledgments}

\bigskip 

\end{document}